\documentstyle[amstex]{article}

\catcode`\@=11
\def\numberbysection{\@addtoreset{equation}{section}
        \def\theequation{\thesection.\arabic{equation}}}
\numberbysection
\input{tcilatex}

\begin{document}

\newlength{\lno} \lno1.5cm \newlength{\len} \len=\textwidth%
\addtolength{\len}{-\lno}

\setcounter{page}{0}

\baselineskip7mm \renewcommand{\thefootnote}{\fnsymbol{footnote}} \newpage %
\setcounter{page}{0}

\begin{titlepage}     
\vspace{0.5cm}
\begin{center}
{\Large\bf  $A_{2}^{(2) }$ Gaudin model and its associated Knizhnik-Zamolodchikov equation}\\
\vspace{1cm}
{\large  V. Kurak $^{\dag }$\hspace{.5cm} and \hspace{.5cm} A. Lima-Santos$^{\ddag}$ } \\
\vspace{1cm}
$^{\dag}${\large \em Universidade de S\~ao Paulo, Instituto de F\'{\i}sica \\
Caixa Postal 66318, CEP 05315-970~~S\~ao Paulo -SP, Brasil}\\
\vspace{.5cm}
$^{\ddag}${\large \em Universidade Federal de S\~ao Carlos, Departamento de F\'{\i}sica \\
Caixa Postal 676, CEP 13569-905~~S\~ao Carlos, Brasil}\\
\end{center}
\vspace{1.2cm}
\begin{abstract}
The semiclassical limit of the algebraic Bethe Ansatz  for the Izergin-Korepin 19-vertex model  is used to solve the 
theory of Gaudin models associated with the twisted $A_{2}^{(2)}$ R-matrix. We find the spectra
and eigenvectors of the $N-1$ independents Gaudin Hamiltonians .
We also use the off-shell Bethe Ansatz method to show how the off-shell Gaudin equation 
solves the associated trigonometric system of Knizhnik-Zamolodchikov equations.
\end{abstract}
\vspace{2cm}
\begin{center}
\end{center}
\vfill
\begin{center}
\small{\today}
\end{center}
\end{titlepage}

\baselineskip6mm

\newpage{}

\section{Introduction}

Most classical statistical mechanics \cite{Baxter, KBI} and two-dimensional
quantum field theory models \cite{Abdalla} exact solubility relies on the $%
{\cal R}$-matrix ${\cal R}(u)$, where $u$ is a spectral parameter, acting on
the tensor product $V\otimes V$ of a given vector space $V$ . ${\cal R}(u)$
abridges the Boltzman weights of classical statistical mechanics models or
the two-body scattering matrix of integrable quantum field theories and is a
solution of the Yang-Baxter ({\small YB}) equation 
\begin{equation}
{\cal R}_{12}(u){\cal R}_{13}(u+v){\cal R}_{23}(v)={\cal R}_{23}(v){\cal R}%
_{13}(u+v){\cal R}_{12}(u),  \label{int.1}
\end{equation}%
in $V^{1}\otimes V^{2}\otimes V^{3}$, where ${\cal R}_{12}={\cal R}\otimes 1$%
, ${\cal R}_{23}=1\otimes {\cal R}$, etc. ${\cal R}(u)$ may also depend on
an additional parameter $\eta $ in such a way that

\begin{equation}
{\cal R}(u,\eta )=1+\eta \ r(u)+{\rm o}(\eta ^{2}),  \label{int.1b}
\end{equation}%
where $1$ is the identity operator in the space $V\otimes V$. In this case
the solution ${\cal R}(u)$ of Eq.(\ref{int.1}) is said to be semiclassical \
and the \textquotedblleft classical $r$-matrix\textquotedblright\ obeys the
equation 
\begin{equation}
\lbrack r_{12}(u),r_{13}(u+v)+r_{23}(v)]+[r_{13}(u+v),r_{23}(v)]=0.
\label{int.2}
\end{equation}%
called the classical Yang-Baxter equation. Nondegenerate solutions of (\ref%
{int.2}) in the tensor product of two copies of a simple Lie algebra {\rm g}
, $r_{ij}(u)\in {\rm g}_{i}\otimes {\rm g}_{j}$ , $i,j=1,2,3$, were
classified by Belavin and Drinfeld \cite{BD} and its important role in the
theory of classical completely integrable systems can be found, for
instance, in \cite{Semenov}.

In the skew-symmetric case $r_{ji}(-u)+r_{ij}(u)=0$, the classical {\small YB%
} equation embodies the compatibility condition for the system of linear
differential equations 
\begin{equation}
\kappa \frac{\partial \Psi (z_{1},...,z_{N})}{\partial z_{i}}%
=\sum\limits_{j\neq i}r_{ij}(z_{i}-z_{j})\Psi (z_{1},...,z_{N})
\label{int.3}
\end{equation}%
in $N$ complex variables $z_{1},...,z_{N}$ \ for a vector-valued function $%
\Psi (z_{1},...,z_{N})$ with values in the tensor product space $%
V=V^{1}\otimes \cdots \otimes V^{N}$ . $\kappa $ is a yet non specified
coupling constant.

In the rational case \cite{BD}, very simple skew-symmetric solutions are
known: $r(u)={\rm C}_{2}/u$, where ${\rm C}_{2}\in {\rm g}\otimes {\rm g}$
is a symmetric invariant tensor of a finite dimensional Lie algebra ${\rm g}$
acting on a representation space $V$, and the above system of linear
differential equations (\ref{int.3}) is known as Knizhnik-Zamolodchikov (%
{\small KZ}) system of equations for the $N-$point conformal blocks of 
{\small WZW} conformal field theories on the sphere \cite{KZ}. The
trigonometric solutions of Eq.(\ref{int.1}), or quantum group solutions,
were classified by Jimbo \cite{Jimbo}, and in this case the system of
equations (\ref{int.3}) is named generalized\ system of {\small KZ }%
differential equations.

The Gaudin \cite{GA} Hamiltonians $G_{i}$ are related to the classical $r$%
-matrices by $\qquad \qquad $%
\begin{equation}
G_{i}=\sum\limits_{j\neq i}r_{ij}(z_{i}-z_{j})  \label{int4a}
\end{equation}%
and the condition for their commutativity is again the classical {\small YB}
equation (\ref{int.2}).

This interplay of the classical {\small YB} equation with conformal field
theory and the theory of Gaudin Hamiltonians can be understood through the
algebraic Bethe Ansatz, here shortly described in the following way.

Babujian and Flume \cite{BAF} showed that the algebraic Bethe Ansatz \cite%
{FT} for the theory of the Gaudin models \cite{GA} can be gotten from the
the algebraic Bethe Ansatz of a non-homogeneous lattice vertex model, the
appropriately expanded $p-$body Bethe wave vectors rendering, in the
semiclassical limit, solutions to the generalized system of {\small KZ}
equations. For instance, in the rational $su(2)$ example, the algebraic
quantum inverse scattering method \cite{FT} allows one to write the
following equation 
\begin{equation}
\tau (u|z)\Phi (u_{1,\cdots ,}u_{p})=\Lambda (u,u_{1},\cdots ,u_{p}|z)\Phi
(u_{1},\cdots ,u_{p})-\sum_{\alpha =1}^{p}\frac{{\cal F}_{\alpha }\Phi
^{\alpha }}{u-u_{\alpha }}.  \label{int.4}
\end{equation}%
Here $\tau (u|z)$ denotes the transfer matrix of the rational vertex model
in an inhomogeneous lattice acting on an $N$-fold tensor product of $su(2)$
representation spaces. $\Phi ^{\alpha }$ \ meaning $\Phi ^{\alpha }=\Phi
(u_{1},\cdots u_{\alpha -1},u,u_{\alpha +1},...,u_{p})$ ; ${\cal F}_{\alpha
}(u_{1},\cdots ,u_{p}|z)$ and $\Lambda (u,u_{1},\cdots ,u_{p}|z)$ being $c$
number functions. The vanishing of the so-called unwanted terms, ${\cal F}%
_{\alpha }=0$, is enforced in the usual procedure of the algebraic Bethe
Ansatz by finding the parameters $u_{1},...,u_{p}$. In this case the wave
vector $\Phi (u_{1},\cdots ,u_{p})$ becomes an eigenvector of the transfer
matrix with eigenvalue $\Lambda (u,u_{1},\cdots ,u_{p}|z)$. If we keep all
unwanted terms, i.e. ${\cal F}_{\alpha }\neq 0$, then the wave vector $\Phi $
in general satisfies the equation (\ref{int.4}), named in \cite{B} as
off-shell Bethe Ansatz equation ({\small OSBAE}).

There is a neat relationship between the wave vector satisfying the {\small %
OSBAE} (\ref{int.4}) and the vector-valued solutions of the {\small KZ}
equation (\ref{int.3}): the general vector valued solution of the {\small KZ}
equation for an arbitrary simple Lie algebra was found by Schechtman and
Varchenko \cite{SV}. It can be represented as a multiple contour integral

\begin{equation}
\Psi (z_{1},\ldots ,z_{N})=\oint \cdots \oint {\cal X}(u_{1},...,u_{p}|z)%
\phi (u_{1},...,u_{p}|z)du_{1}\cdots du_{p}.  \label{int.5}
\end{equation}%
where the complex variables $z_{1},...,z_{N}$ of (\ref{int.5}) are related
with the disorder parameters of the {\small OSBAE} . The vector valued
function $\phi (u_{1},...,u_{p}|z)$ is the semiclassical limit of the wave
vector $\Phi (u_{1},...,u_{p}|z)$. In fact, it is the Bethe wave vector for
Gaudin magnets \cite{GA}, but \ "off -shell ". The Bethe Ansatz for the
Gaudin model was derived for any simple Lie algebra by Reshetikhin and
Varchenko \cite{RV}. The scalar function ${\cal X}(u_{1},...,u_{p}|z)$ is
constructed from the semiclassical limit of the $\Lambda
(u=z_{k};u_{1},...,u_{p}|z)$ and the ${\cal F}_{\alpha }(u_{1},\cdots
,u_{p}|z)$ functions.

In this paper we investigate the semiclassical limit of Izegin-Korepin (%
{\small IK}) model \cite{IK}, which corresponds to the twisted affine Lie
algebra $A_{2}^{(2)}$ solution of \cite{Jimbo}.

The paper is organized as follows. In Section $2$ we present the algebraic
Bethe Ansatz for the $A_{2}^{(2)}$ vertex model. Here the inhomogeneous
Bethe Ansatz is read from the homogeneous case previously derived for the $%
19 $-vertex models \cite{TA, LI}. We also derive the off-shell Bethe Ansatz
equation for this vertex model. In Section $3$ , taking into account the
semiclassical limit of the results presented in the Section $2$, we describe
the algebraic structure of the corresponding Gaudin model. In Section $4$,
data of the off-shell Gaudin equation are used to construct solutions of the
trigonometric {\small KZ} equation. Conclusions are reserved for Section $5$.

\section{Inhomogeneous Algebraic Bethe Ansatz}

Besides ${\cal R}$ , it is usual to consider the matrix $R={\cal PR}$ that
satisfies the equation 
\begin{equation}
R_{12}(u)R_{23}(u+v)R_{12}(v)=R_{23}(v)R_{12}(u+v)R_{23}(u).  \label{gra.2}
\end{equation}

We choose the regular solution of the {\small YB} equation for the {\small IK%
} $19$-vertex model \cite{IK} to be normalized such that 
\begin{equation}
R(u,\eta )=\left( 
\begin{array}{ccccccccccc}
x_{1} & 0 & 0 &  & 0 & 0 & 0 &  & 0 & 0 & 0 \\ 
0 & y_{5} & 0 &  & x_{2} & 0 & 0 &  & 0 & 0 & 0 \\ 
0 & 0 & y_{7} &  & 0 & y_{6} & 0 &  & x_{3} & 0 & 0 \\ 
&  &  &  &  &  &  &  &  &  &  \\ 
0 & x_{2} & 0 &  & x_{5} & 0 & 0 &  & 0 & 0 & 0 \\ 
0 & 0 & y_{6} &  & 0 & x_{4} & 0 &  & x_{6} & 0 & 0 \\ 
0 & 0 & 0 &  & 0 & 0 & y_{5} &  & 0 & x_{2} & 0 \\ 
&  &  &  &  &  &  &  &  &  &  \\ 
0 & 0 & x_{3} &  & 0 & x_{6} & 0 &  & x_{7} & 0 & 0 \\ 
0 & 0 & 0 &  & 0 & 0 & x_{2} &  & 0 & x_{5} & 0 \\ 
0 & 0 & 0 &  & 0 & 0 & 0 &  & 0 & 0 & x_{1}%
\end{array}%
\right) ,  \label{gra.3}
\end{equation}%
where 
\begin{eqnarray}
x_{1}(u) &=&\sinh (u+2\eta )\cosh (u+3\eta ),\quad x_{2}(u)=\sinh u\cosh
(u+3\eta ),  \nonumber \\
x_{3}(u) &=&\sinh u\cosh (u+\eta ),\quad x_{4}(u)=\sinh u\cosh (u+3\eta
)+\sinh 2\eta \cosh 3\eta ,\quad  \nonumber \\
x_{5}(u) &=&{\rm e}^{u}\sinh 2\eta \cosh (u+3\eta ),\quad y_{5}(u)={\rm e}%
^{-u}\sinh 2\eta \cosh (u+3\eta ),\qquad  \nonumber \\
x_{6}(u) &=&{\rm e}^{u+2\eta }\sinh 2\eta \sinh u,\quad y_{6}(u)\ =-{\rm e}%
^{-u-2\eta }\sinh 2\eta \sinh u,\qquad  \nonumber \\
x_{7}(u) &=&{\rm e}^{2u}x_{1}(u)-{\rm e}^{2u+4\eta }x_{3}(u),\qquad y_{7}(u)=%
{\rm e}^{-2u}x_{1}(u)-{\rm e}^{-2u-4\eta }x_{3}(u)  \label{gra.4}
\end{eqnarray}

An inhomogeneous vertex model has two parameters: a global spectral
parameter $u$ and a disorder parameter $z$, so that the vertex weight matrix 
${\cal R}$ depends on the difference $u-z$ and consequently the monodromy
matrix defined bellow will be a function of the disorder parameters $z_{i}$.

The quantum inverse scattering method is characterized by the monodromy
matrix $T(u|z)$ satisfying the equation 
\begin{equation}
R(u-v)\left[ T(u|z)\otimes T(v|z)\right] =\left[ T(v|z)\otimes T(u|z)\right]
R(u-v),  \label{gra.5}
\end{equation}%
whose existence is guaranteed by the {\small YB} equation (\ref{int.1}). $%
T(u|z)$ is a matrix in the space $V$ (usually called auxiliary space) whose
matrix elements are operators on the states of the quantum system (which
will also be the space $V$ in this work ). The monodromy operator $T(u|z)$
is defined as an ordered product of local operators ${\cal L}_{n}$ (Lax
operator), on all sites of the lattice: 
\begin{equation}
T(u|z)={\cal L}_{N}(u-z_{N}){\cal L}_{N-1}(u-z_{N-1})\cdots {\cal L}%
_{1}(u-z_{1}).  \label{gra.6}
\end{equation}%
We normalize the Lax operator on the $n^{th}$ quantum space to be given by: 
\begin{eqnarray}
{\cal L}_{n} &=&\frac{1}{x_{2}}\left( 
\begin{array}{ccccccccccc}
x_{1} & 0 & 0 &  & 0 & 0 & 0 &  & 0 & 0 & 0 \\ 
0 & x_{2} & 0 &  & x_{5} & 0 & 0 &  & 0 & 0 & 0 \\ 
0 & 0 & x_{3} &  & 0 & x_{6} & 0 &  & x_{7} & 0 & 0 \\ 
&  &  &  &  &  &  &  &  &  &  \\ 
0 & y_{5} & 0 &  & x_{2} & 0 & 0 &  & 0 & 0 & 0 \\ 
0 & 0 & y_{6} &  & 0 & x_{4} & 0 &  & x_{6} & 0 & 0 \\ 
0 & 0 & 0 &  & 0 & 0 & x_{2} &  & 0 & x_{5} & 0 \\ 
&  &  &  &  &  &  &  &  &  &  \\ 
0 & 0 & x_{7} &  & 0 & y_{6} & 0 &  & x_{3} & 0 & 0 \\ 
0 & 0 & 0 &  & 0 & 0 & y_{5} &  & 0 & x_{2} & 0 \\ 
0 & 0 & 0 &  & 0 & 0 & 0 &  & 0 & 0 & x_{1}%
\end{array}%
\right)  \nonumber \\
&=&\left( 
\begin{array}{lll}
L_{11}^{(n)}(u-z_{n}) & L_{12}^{(n)}(u-z_{n}) & L_{13}^{(n)}(u-z_{n}) \\ 
L_{21}^{(n)}(u-z_{n}) & L_{22}^{(n)}(u-z_{n}) & L_{23}^{(n)}(u-z_{n}) \\ 
L_{31}^{(n)}(u-z_{n}) & L_{32}^{(n)}(u-z_{n}) & L_{33}^{(n)}(u-z_{n})%
\end{array}%
\right)  \label{gra.7}
\end{eqnarray}%
$L_{\alpha \beta }^{(n)}(u),\ \alpha ,\beta =1,2,3$ are $3$ by $3$ matrices
acting on the $n^{th}$ site of the lattice, so the monodromy matrix has the
form 
\begin{equation}
T(u|z)=\left( 
\begin{array}{lll}
A_{1}(u|z) & B_{1}(u|z) & B_{2}(u|z) \\ 
C_{1}(u|z) & A_{2}(u|z) & B_{3}(u|z) \\ 
C_{2}(u|z) & C_{3}(u|z) & A_{3}(u|z)%
\end{array}%
\right) ,  \label{gra.8}
\end{equation}%
where 
\begin{eqnarray}
T_{ij}(u|z)
&=&\sum_{k_{1},...,k_{N-1}=1}^{3}L_{ik_{1}}^{(N)}(u-z_{N})\otimes
L_{k_{1}k_{2}}^{(N-1)}(u-z_{N-1})\otimes \cdots \otimes
L_{k_{N-1}j}^{(1)}(u-z_{1}).  \nonumber \\
i,j &=&1,2,3.  \label{gra.9}
\end{eqnarray}

The vector in the quantum space of the monodromy matrix $T(u|z)$ that is
annihilated by the operators $T_{ij}(u|z)$, $i>j$ ($C_{i}(u|z)$ operators, $%
i=1,2,3$) and it is also an eigenvector for the operators $T_{ii}(u|z)$ ( $%
A_{i}(u|z)$ operators, $i=1,2,3$) is called a highest \ weight vector of the
monodromy matrix $T(u|z)$.

The transfer matrix $\tau (u|z)$ of the corresponding integrable spin model
is given by the trace of the monodromy matrix in the space $V$ 
\begin{equation}
\tau (u|z)=A_{1}(u|z)+A_{2}(u|z)+A_{3}(u|z).  \label{gra.10}
\end{equation}

The algebraic Bethe Ansatz solution for the inhomogeneous {\small IK} vertex
model can be obtained from the homogeneous case \cite{TA}. The proper
modification is a local shift of the spectral parameter $u\rightarrow
u-z_{i} $, and the functions needed in the sequel are

\begin{eqnarray}
z(u) &=&\frac{x_{1}(u)}{x_{2}(u)}=\frac{\sinh (u+2\eta )}{\sinh u},\quad
y(u)=\frac{x_{3}(u)}{y_{6}(u)}=-{\rm e}^{u+2\eta }\frac{\cosh (u+\eta )}{%
\sinh 2\eta },\quad   \nonumber \\
&&  \nonumber \\
\omega (u) &=&\frac{x_{1}(u)x_{3}(u)}{x_{4}(u)x_{3}(u)-x_{6}(u)y_{6}(u)}=%
\frac{\cosh (u+\eta )}{\cosh (u-\eta )},  \nonumber \\
&&  \nonumber \\
{\cal Z}(u_{k}-u_{j}) &=&\left\{ 
\begin{array}{c}
z(u_{k}-u_{j})\qquad \qquad \quad \quad {\rm if}\quad k>j \\ 
z(u_{k}-u_{j})\omega (u_{j}-u_{k})\quad \ {\rm if}\quad k<j%
\end{array}%
\right. .  \label{inh.1}
\end{eqnarray}%
We start defining the highest weight vector of the monodromy matrix $T(u|z)$
in a lattice of $N$ sites as the completely unoccupied state 
\begin{equation}
\left\vert 0\right\rangle =\otimes _{a=1}^{N}\left( 
\begin{array}{c}
1 \\ 
0 \\ 
0%
\end{array}%
\right) _{a}.  \label{inh.2}
\end{equation}%
Using (\ref{gra.9}) we can compute the normalized action of the monodromy
matrix entries on this state 
\begin{eqnarray}
A_{i}(u|z)\left\vert 0\right\rangle  &=&X_{i}(u|z)\left\vert 0\right\rangle
,\quad C_{i}(u|z)\left\vert 0\right\rangle =0,\quad B_{i}(u|z)\left\vert
0\right\rangle \neq \left\{ 0,\left\vert 0\right\rangle \right\} ,  \nonumber
\\
X_{i}(u|z) &=&\prod_{a=1}^{N}\frac{x_{i}(u-z_{a})}{x_{2}(u-z_{a})},\qquad
i=1,2,3.  \label{inh.3}
\end{eqnarray}

The Bethe vectors are defined as normal ordered states $\Psi
_{n}(u_{1},\cdots ,u_{n})$ which can be written with aid of the recurrence
formula \cite{TA}: 
\begin{eqnarray}
&&\left. \Psi _{n}(u_{1},...,u_{n}|z)=B_{1}(u_{1}|z)\Psi
_{n-1}(u_{2},...,u_{n}|z)\right.   \nonumber \\
&&  \nonumber \\
&&\left. -B_{2}(u_{1}|z)\sum_{j=2}^{n}\frac{X_{1}(u_{j}|z)}{y(u_{1}-u_{j})}%
\prod_{k=2,k\neq j}^{n}{\cal Z}(u_{k}-u_{j})\Psi _{n-2}(u_{2},...,\overset{%
\wedge }{u}_{j},...,u_{n}|z)\right. ,  \label{inh.5}
\end{eqnarray}%
with the initial condition $\Psi _{0}=\left\vert 0\right\rangle ,\quad \Psi
_{1}(u_{1}|z)=B_{1}(u_{1}|z)\left\vert 0\right\rangle $. Here \ $\overset{%
\wedge }{u}_{j}$ denotes that the parameter $u_{j}$ is absent: $\Psi (%
\overset{\wedge }{u}_{j}|z)=\Psi (u_{1},...,u_{j-1},u_{j+1},\cdots ,u_{n}|z)$%
.

The action of the transfer matrix $\tau (u|z)$ on the Bethe vectors gives us
the following off-shell Bethe Ansatz equation for the $A_{2}^{(2)}$ vertex
model

\begin{equation}
\tau (u|z)\Psi _{n}(u_{1},...,u_{n}|z)=\Lambda _{n}\Psi
_{n}(u_{1},...,u_{n}|z)-\sum_{j=1}^{n}{\cal F}_{j}^{(n-1)}\Psi
_{(n-1)}^{j}+\sum_{j=2}^{n}\sum_{l=1}^{j-1}{\cal F}_{lj}^{(n-2)}\Psi
_{(n-2)}^{lj}.  \label{inh.4}
\end{equation}

We now briefly describe each term which appear in the right hand side of (%
\ref{inh.4}) ( for more details the reader can see \cite{LI}): in the first
term the Bethe vectors (\ref{inh.5}) are multiplied by $c$-number functions $%
\Lambda _{n}=\Lambda _{n}(u,u_{1},...,u_{n}|z)$ given by 
\begin{equation}
\Lambda _{n}=X_{1}(u|z)\prod_{k=1}^{n}z(u_{k}-u)+X_{2}(u|z)\prod_{k=1}^{n}%
\frac{z(u-u_{k})}{\omega (u-u_{k})}+X_{3}(u|z)\prod_{k=1}^{n}\frac{%
x_{2}(u-u_{k})}{x_{3}(u-u_{k})}.  \label{inh.6}
\end{equation}%
The second term is a sum of new vectors 
\begin{equation}
\Psi _{(n-1)}^{j}=\left( \frac{x_{5}(u_{j}-u)}{x_{2}(u_{j}-u)}B_{1}(u|z)-%
\frac{1}{y(u-u_{j})}B_{3}(u|z)\right) \Psi _{n-1}(\overset{\wedge }{u}_{j}),
\label{inh.7}
\end{equation}%
multiplied by $c-$number functions ${\cal F}_{j}^{(n-1)}$ given by 
\begin{equation}
{\cal F}_{j}^{(n-1)}=X_{1}(u_{j}|z)\prod_{k\neq j}^{n}{\cal Z}%
(u_{k}-u_{j})-X_{2}(u_{j}|z)\prod_{k\neq j}^{n}{\cal Z}(u_{j}-u_{k}).
\label{inh.8}
\end{equation}%
Finally, the last term is a coupled sum of a third type of vector-valued
functions 
\begin{equation}
\Psi _{(n-2)}^{lj}=B_{2}(u|z)\Psi _{n-2}(\overset{\wedge }{u}_{l},\overset{%
\wedge }{u}_{j}),  \label{inh.9}
\end{equation}%
with $c-$number coefficients 
\begin{eqnarray}
{\cal F}_{lj}^{(n-2)} &=&G_{lj}X_{1}(u_{l}|z)X_{1}(u_{j}|z)\prod_{k=1,k\neq
j,l}^{n}{\cal Z}(u_{k}-u_{l}){\cal Z}(u_{k}-u_{j})  \nonumber \\
&&+Y_{lj}X_{1}(u_{l}|z)X_{2}(u_{j}|z)\prod_{k=1,k\neq j,l}^{n}{\cal Z}%
(u_{k}-u_{l}){\cal Z}(u_{j}-u_{k})  \nonumber \\
&&+F_{lj}X_{1}(u_{j}|z)X_{2}(u_{l}|z)\prod_{k=1,k\neq j,l}^{n}{\cal Z}%
(u_{l}-u_{k}){\cal Z}(u_{k}-u_{j})  \nonumber \\
&&+H_{lj}X_{2}(u_{l}|z)X_{2}(u_{j}|z)\prod_{k=1,k\neq j,l}^{n}{\cal Z}%
(u_{j}-u_{k}){\cal Z}(u_{l}-u_{k}).  \label{inh.10}
\end{eqnarray}%
where $G_{lj}$ , $Y_{lj}$ , $F_{lj}$ and $H_{lj}$ are additional functions
defined by 
\begin{eqnarray}
G_{lj} &=&\frac{x_{7}(u_{l}-u)}{x_{3}(u_{l}-u)}\frac{1}{y(u_{l}-u_{j})}+%
\frac{z(u_{l}-u)}{\omega (u_{l}-u)}\frac{x_{5}(u_{j}-u)}{x_{2}(u_{j}-u)}%
\frac{1}{y(u-u_{l})},  \nonumber \\
&&  \nonumber \\
H_{lj} &=&\frac{y_{7}(u-u_{l})}{x_{3}(u-u_{l})}\frac{1}{y(u_{l}-u_{j})}-%
\frac{y_{5}(u-u_{l})}{x_{3}(u-u_{l})}\frac{1}{y(u-u_{j})},  \nonumber \\
&&  \nonumber \\
Y_{lj} &=&\frac{1}{y(u-u_{l})}\left\{ z(u-u_{l})\frac{y_{5}(u-u_{j})}{%
x_{2}(u-u_{j})}-\frac{y_{5}(u-u_{l})}{x_{2}(u-u_{l})}\frac{y_{5}(u_{l}-u_{j})%
}{x_{2}(u_{l}-u_{j})}\right\} ,  \nonumber \\
&&  \nonumber \\
F_{lj} &=&\frac{y_{5}(u-u_{l})}{x_{2}(u-u_{l})}\left\{ \frac{%
y_{5}(u_{l}-u_{j})}{x_{2}(u_{l}-u_{j})}\frac{1}{y(u-u_{l})}+\frac{z(u-u_{l})%
}{\omega (u-u_{l})}\frac{1}{y(u-u_{j})}\right.  \nonumber \\
&&\left. -\frac{x_{5}(u-u_{l})}{x_{2}(u-u_{l})}\frac{1}{y(u_{l}-u_{j})}%
\right\} .  \label{inh.11}
\end{eqnarray}%
In the usual Bethe Ansatz method, the next step is to be impose the
vanishing of the so-called unwanted terms of (\ref{inh.4}) in order to get
an eigenvalue problem for the transfer matrix.

We impose ${\cal F}_{j}^{(n-1)}=0$ and ${\cal F}_{lj}^{(n-2)}=0$ into (\ref%
{inh.4}) to recover the eigenvalue problem. This means that $\Psi
_{n}(u_{1},...,u_{n}|z)$ is an eigenstate of $\tau (u|z)$ with eigenvalue $%
\Lambda _{n}$ , provided the parameters $u_{j}$ are solutions of the
inhomogeneous Bethe Ansatz equations 
\begin{eqnarray}
\prod_{a=1}^{N}z(u_{j}-z_{a}) &=&\prod_{k=1,\ k\neq j}^{n}\frac{%
z(u_{j}-u_{k})}{z(u_{k}-u_{j})}\omega (u_{k}-u_{j}),  \nonumber \\
j &=&1,2,...,n.  \label{inh.12}
\end{eqnarray}

\newpage

\section{Structure of the ${\bf A}_{2}^{(2)}$ Gaudin Model}

In this section we will consider the theory of the Gaudin models and to do
that we need to compute the expansions of some results presented in the
previous section in a power series of $\eta $ .

As we already mentioned the IK vertex model corresponds , in the Jimbo \cite%
{Jimbo} quantum group classification, to the twisted affine Lie algebra $%
A_{2}^{(2)}$,which is constructed out of the second order automorphism of
the Lie algebra $A_{2}$ , the swapping of its simple roots, such that the
fixed algebra under this automorphism, say $g_{0}$, is isomorphic to the Lie
algebra $A_{1}$. We recall \cite{Corn} the following base of \ "twisted " $%
su(3)$ generators in the fundamental representation 
\begin{eqnarray}
H &=&\left( 
\begin{array}{lll}
1 & 0 & 0 \\ 
0 & 0 & 0 \\ 
0 & 0 & -1%
\end{array}%
\right) ,\quad U^{+}=\left( 
\begin{array}{lll}
0 & 1 & 0 \\ 
0 & 0 & 1 \\ 
0 & 0 & 0%
\end{array}%
\right) ,\ \quad U^{-}=\left( 
\begin{array}{lll}
0 & 0 & 0 \\ 
1 & 0 & 0 \\ 
0 & 1 & 0%
\end{array}%
\right) ,\qquad  \nonumber \\
&&  \nonumber \\
\ S^{+} &=&\left( 
\begin{array}{lll}
0 & 0 & 1 \\ 
0 & 0 & 0 \\ 
0 & 0 & 0%
\end{array}%
\right) ,\quad S^{-}=\left( 
\begin{array}{lll}
0 & 0 & 0 \\ 
0 & 0 & 0 \\ 
1 & 0 & 0%
\end{array}%
\right) ,\ \quad Y=\left( 
\begin{array}{lll}
1 & 0 & 0 \\ 
0 & -2 & 0 \\ 
0 & 0 & 1%
\end{array}%
\right) ,  \nonumber \\
&&  \nonumber \\
W^{+} &=&\left( 
\begin{array}{lll}
0 & 1 & 0 \\ 
0 & 0 & -1 \\ 
0 & 0 & 0%
\end{array}%
\right) ,\ W^{-}=\left( 
\begin{array}{lll}
0 & 0 & 0 \\ 
1 & 0 & 0 \\ 
0 & -1 & 0%
\end{array}%
\right)  \label{str.1}
\end{eqnarray}%
$H$, $U^{+}$ and $U^{-}$ correspond to the even sector of the automorphism
and generate $g_{0}$ , the remaining generators correspond to the odd sector
and transform according to the \ "isospin " $2$ \ representation of \ $g_{0}$%
.

The quadratic Casimir operator is 
\begin{equation}
C_{2}=H^{2}+\frac{1}{3}Y^{2}+2\{S^{+},S^{-}\}+\{U^{+},U^{-}\}+\{W^{+},W^{-}%
\}.  \label{str.2}
\end{equation}%
where $\{A,B\}=AB+BA$.

In order to expand the matrix elements of $T(u|z)$, up to an appropriate
order in $\eta $, we will start by expanding the Lax operator entries
defined in (\ref{gra.7}): 
\begin{eqnarray}
L_{11}^{(n)} &=&1+2\eta \frac{\frac{1}{3}(2+Y_{n})+H_{n}\cosh (2u-2z_{n})}{%
\sinh (2u-2z_{n})}+4\eta ^{2}\left( \frac{1}{2}H_{n}^{2}-\frac{3}{4}\frac{%
H_{n}^{2}-H_{n}}{\cosh (u-z_{n})^{2}}\right) +{\rm o}(\eta ^{3}),  \nonumber
\\
&&  \nonumber \\
L_{22}^{(n)} &=&1+2\eta \frac{\frac{2}{3}(1-Y_{n})}{\sinh (2u-2z_{n})}-4\eta
^{2}\left( \frac{3}{4}\frac{H_{n}^{2}-Y_{n}}{\cosh (u-z_{n})^{2}}\right) +%
{\rm o}(\eta ^{3}),  \nonumber \\
&&  \nonumber \\
L_{33}^{(n)} &=&1+2\eta \frac{\frac{1}{3}(2+Y_{n})-H_{n}\cosh (2u-2z_{n})}{%
\sinh (2u-2z_{n})}+4\eta ^{2}\left( \frac{1}{2}H_{n}^{2}-\frac{3}{4}\frac{%
H_{n}^{2}+H_{n}}{\cosh (u-z_{n})^{2}}\right) +{\rm o}(\eta ^{3}).  \nonumber
\\
&&  \label{gau.1}
\end{eqnarray}%
and for the elements out of the diagonal we have 
\begin{eqnarray}
L_{12}^{(n)} &=&2\eta \frac{W_{a}^{-}+{\rm e}^{2(u-z_{n})}U_{a}^{-}}{\sinh
(2u-2z_{n})}+{\rm o}(\eta ^{2}),\quad \ L_{21}^{(n)}=2\eta \frac{W_{a}^{+}+%
{\rm e}^{-2(u-z_{n})}U_{a}^{+}}{\sinh (2u-2z_{n})}+{\rm o}(\eta ^{2}), 
\nonumber \\
&&  \nonumber \\
L_{23}^{(n)} &=&-2\eta \frac{W_{a}^{-}-{\rm e}^{2(u-z_{n})}U_{a}^{-}}{\sinh
(2u-2z_{n})}+{\rm o}(\eta ^{2}),\quad \quad L_{32}^{(n)}=-2\eta \frac{%
W_{a}^{+}-{\rm e}^{-2(u-z_{n})}U_{a}^{+}}{\sinh (2u-2z_{n})}+{\rm o}(\eta
^{2}),  \nonumber \\
&&  \nonumber \\
L_{13}^{(n)} &=&2\eta \ \frac{2S_{n}^{-}}{\sinh (2u-2z_{n})}+{\rm o}(\eta
^{2}),\quad \quad L_{31}^{(n)}=2\eta \ \frac{2S_{n}^{+}}{\sinh (2u-2z_{n})}+%
{\rm o}(\eta ^{2}).  \label{gau.2}
\end{eqnarray}

Substituting (\ref{gau.1}) and (\ref{gau.2}) into (\ref{gra.9}) we will get
the semiclassical expansion for the monodromy matrix entries.\ For the
diagonal entries we get 
\begin{equation}
A_{i}(u|z)=1+2\eta A_{i}^{(1)}(u|z)+4\eta ^{2}A_{i}^{(2)}(u|z)+{\rm o}(\eta
^{3}),\quad i=1,2,3.
\end{equation}%
where, the first order terms are given by%
\begin{eqnarray}
A_{1}^{(1)}(u|z) &=&\sum_{a=1}^{N}\frac{\frac{1}{3}(2+Y_{a})+H_{a}\cosh
(2u-2z_{a})}{\sinh (2u-2z_{a})},\quad A_{2}^{(1)}(u|z)=\sum_{a=1}^{N}\frac{%
\frac{2}{3}(1-Y_{a})}{\sinh (2u-2z_{a})},  \nonumber \\
A_{3}^{(1)}(u|z) &=&\sum_{a=1}^{N}\frac{\frac{1}{3}(2+Y_{a})-H_{a}\cosh
(2u-2z_{a})}{\sinh (2u-2z_{a})}
\end{eqnarray}%
and the second order terms are%
\begin{eqnarray*}
A_{1}^{(2)}(u|z) &=&\sum_{a=1}^{N}(\frac{1}{2}H_{a}^{2}-\frac{3}{4}\frac{%
H_{a}^{2}-H_{a}}{\cosh (2u-2z_{a})^{2}})+\sum_{a<b}\coth (2u-2z_{a})\coth
(2u-2z_{b})H_{a}\otimes H_{b} \\
&&+\sum_{a<b}\frac{\frac{1}{3}(2+Y_{a})\otimes \frac{1}{3}(2+Y_{b})}{\sinh
(2u-2z_{a})\sinh (2u-2z_{b})}+\sum_{a<b}\frac{\frac{1}{3}(2+Y_{a})\otimes
H_{b}\cosh (2u-2z_{b})}{\sinh (2u-2z_{a})\sinh (2u-2z_{b})} \\
&&+\sum_{a<b}\frac{\cosh (2u-2z_{a})H_{a}\otimes \frac{1}{3}(2+Y_{b})}{\sinh
(2u-2z_{a})\sinh (2u-2z_{b})}+\sum_{a<b}4\frac{S_{a}^{-}\otimes S_{b}^{+}}{%
\sinh (2u-2z_{a})\sinh (2u-2z_{b})} \\
&&+\sum_{a<b}\frac{W_{a}^{-}+{\rm e}^{2(u-z_{a})}U_{a}^{-}}{\sinh (2u-2z_{a})%
}\otimes \frac{W_{b}^{+}+{\rm e}^{-2(u-z_{b})}U_{b}^{+}}{\sinh (2u-2z_{b})},
\end{eqnarray*}%
\begin{eqnarray*}
A_{3}^{(2)}(u|z) &=&\sum_{a=1}^{N}(\frac{1}{2}H_{a}^{2}-\frac{3}{4}\frac{%
H_{a}^{2}+H_{a}}{\cosh (2u-2z_{a})^{2}})+\sum_{a<b}\coth (2u-2z_{a})\coth
(2u-2z_{b})H_{a}\otimes H_{b} \\
&&+\sum_{a<b}\frac{\frac{1}{3}(2+Y_{a})\otimes \frac{1}{3}(2+Y_{b})}{\sinh
(2u-2z_{a})\sinh (2u-2z_{b})}-\sum_{a<b}\frac{\frac{1}{3}(2+Y_{a})\otimes
H_{b}\cosh (2u-2z_{b})}{\sinh (2u-2z_{a})\sinh (2u-2z_{b})} \\
&&-\sum_{a<b}\frac{\cosh (2u-2z_{a})H_{a}\otimes \frac{1}{3}(2+Y_{b})}{\sinh
(2u-2z_{a})\sinh (2u-2z_{b})}+\sum_{a<b}4\frac{S_{a}^{+}\otimes S_{b}^{-}}{%
\sinh (2u-2z_{a})\sinh (2u-2z_{b})} \\
&&+\sum_{a<b}\left( \frac{W_{a}^{+}-{\rm e}^{-2(u-z_{a})}U_{a}^{+}}{\sinh
(2u-2z_{a})}\right) \otimes \left( \frac{W_{b}^{-}-{\rm e}%
^{2(u-z_{b})}U_{b}^{-}}{\sinh (2u-2z_{b})}\right) ,
\end{eqnarray*}%
\begin{eqnarray}
A_{2}^{(2)}(u|z) &=&\sum_{a=1}^{N}\left( \frac{3}{4}\frac{Y_{a}-H_{a}^{2}}{%
\cosh (2u-2z_{a})^{2}}\right) +\sum_{a<b}\frac{\frac{2}{3}(1-Y_{a})\otimes 
\frac{2}{3}(1-Y_{b})}{\sinh (2u-2z_{a})\sinh (2u-2z_{b})}  \nonumber \\
&&+\sum_{a<b}\frac{W_{a}^{+}+{\rm e}^{-2(u-z_{a})}U_{a}^{+}}{\sinh
(2u-2z_{a})}\otimes \frac{W_{b}^{-}+{\rm e}^{2(u-z_{b})}U_{b}^{-}}{\sinh
(2u-2z_{b})}  \nonumber \\
&&+\sum_{a<b}\frac{W_{a}^{-}-{\rm e}^{2(u-z_{a})}U_{a}^{-}}{\sinh (2u-2z_{a})%
}\otimes \frac{W_{b}^{+}-{\rm e}^{-2(u-z_{b})}U_{b}^{+}}{\sinh (2u-2z_{b})}.
\label{gau.3}
\end{eqnarray}%
It will be only needed to expand the off-diagonal elements up to the first
order in $\eta $%
\begin{eqnarray}
B_{1}(u|z) &=&2\eta \sum_{a=1}^{N}\frac{W_{a}^{-}+{\rm e}%
^{2(u-z_{a})}U_{a}^{-}}{\sinh (2u-2z_{a})}+{\rm o}(\eta ^{2}),\quad 
\nonumber \\
B_{2}(u|z) &=&2\eta \sum_{a=1}^{N}\frac{2S_{a}^{-}}{\sinh (2u-2z_{a})}+{\rm o%
}(\eta ^{2}),  \nonumber \\
B_{3}(u|z) &=&-2\eta \sum_{a=1}^{N}\frac{W_{a}^{-}-{\rm e}%
^{2(u-z_{a})}U_{a}^{-}}{\sinh (2u-2z_{a})}+{\rm o}(\eta ^{2}),  \nonumber \\
C_{1}(u|z) &=&2\eta \sum_{a=1}^{N}\frac{W_{a}^{+}+{\rm e}%
^{-2(u-z_{a})}U_{a}^{+}}{\sinh (2u-2z_{a})}+{\rm o}(\eta ^{2}),  \nonumber \\
C_{2}(u|z) &=&2\eta \sum_{a=1}^{N}\frac{2S_{a}^{+}}{\sinh (u-z_{a})}+{\rm o}%
(\eta ^{2})  \nonumber \\
C_{3}(u|z) &=&-2\eta \sum_{a=1}^{N}\frac{W_{a}^{+}-{\rm e}%
^{-2(u-z_{a})}U_{a}^{+}}{\sinh (2u-2z_{a})}+{\rm o}(\eta ^{2}),
\label{gau.4}
\end{eqnarray}%
>From these expansions we have the following expansion for the transfer
matrix (\ref{gra.10}): 
\begin{eqnarray}
\tau (u|z) &=&3+2\eta \sum_{a=1}^{N}\frac{2}{\sinh (2u-2z_{a})}+4\eta
^{2}\left\{ \sum_{a=1}^{N}(H_{a}^{2}-\frac{3}{2}\frac{1}{\cosh (u-z_{a})^{2}}%
)\right.  \nonumber \\
&&+\sum_{a<b}^{N}\frac{2}{\sinh (2u-2z_{a})\sinh (2u-2z_{b})}\!\!\left\{ 
\frac{2}{3}+H_{a}\overset{s}{\otimes }H_{b}\cosh (2u-2z_{a})\cosh
(2u-2z_{b})\right.  \nonumber \\
&&+\left. \frac{1}{3}Y_{a}\otimes Y_{b}+2(S_{a}^{+}\otimes
S_{b}^{-}+S_{a}^{-}\otimes S_{b}^{+})\right. +W_{a}^{+}\otimes
W_{b}^{-}+W_{a}^{-}\otimes W_{b}^{+}  \nonumber \\
&&+\left. {\rm e}^{2z_{a}-2z_{b}}U_{a}^{+}\otimes U_{b}^{-}+{\rm e}%
^{-2z_{a}+2z_{b}}U_{a}^{-}\otimes U_{b}^{+}\!\!\right\}  \nonumber \\
&\equiv &3+2\eta \tau ^{(1)}(u|z)+4\eta ^{2}\tau ^{(2)}(u|z)+{\rm o}(\eta
^{2}).  \label{gau.5}
\end{eqnarray}%
The second order contribution for $\tau (u|z)$ is%
\begin{equation}
\tau ^{(2)}(u|z)=\sum_{a<b}^{N}{\cal G}_{ab}(u)+\sum_{a=1}^{N}(H_{a}^{2}-%
\frac{3}{2}\frac{1}{\cosh (u-z_{a})^{2}}).  \label{gau.5a}
\end{equation}%
which, with aid of the identity%
\begin{equation}
\frac{1}{\sinh (2u-2z_{a})\sinh (2u-2z_{b})}=\frac{1}{\sinh (2z_{a}-2z_{b})}(%
\frac{{\rm e}^{-2(u-z_{a})}}{\sinh (2u-2z_{a})}-\frac{{\rm e}^{-2(u-2z_{b})}%
}{\sinh (2u-2z_{b})})  \label{gau.5b}
\end{equation}%
can be written in the form%
\begin{equation}
\tau ^{(2)}(u|z)=\sum_{a=1}^{N}\frac{2{\cal G}_{a}(u)}{{\rm e}%
^{2(u-z_{a})}\sinh (2u-2z_{a})}+\sum_{a=1}^{N}(H_{a}^{2}-\frac{3}{2}\frac{1}{%
\cosh (u-z_{a})^{2}})  \label{gau.5c}
\end{equation}%
where 
\begin{eqnarray}
{\cal G}_{a}(u) &=&\sum_{b\neq a}\frac{1}{\sinh (2z_{a}-2z_{b})}\left\{ 
\frac{2}{3}+\cosh (2u-2z_{a})\cosh (2u-2z_{b})H_{a}\otimes H_{b}\right. 
\nonumber \\
&&+\left. \frac{1}{3}Y_{a}\otimes Y_{b}+2\left( S_{a}^{+}\otimes
S_{b}^{-}+S_{a}^{-}\otimes S_{b}^{+}\right) +W_{a}^{-}\otimes
W_{b}^{+}+W_{a}^{+}\otimes W_{b}^{-}\right.  \nonumber \\
&&\left. {\rm e}^{2(z_{a}-z_{b})}U_{a}^{+}\otimes U_{b}^{-}+{\rm e}%
^{-2(z_{a}-z_{b})}U_{a}^{-}\otimes U_{b}^{+}\!\!\right\}  \nonumber \\
&&  \label{gau.5d}
\end{eqnarray}%
Here we observe that \ ${\cal G}_{a}(u)$ is nothing but the sum of
semiclassical \ trigonometric $r$-matrices. This fact follows from the
construction of the semiclassical $r$-matrices out of the quadratic Casimir:%
\begin{equation}
C_{2}=H^{2}+\frac{1}{3}%
Y^{2}+2(S^{-}S^{+}+S^{+}S^{-})+(U^{-}U^{+}+U^{+}U^{-})+(W^{+}W^{-}+W^{-}W^{+}).
\label{gau.6}
\end{equation}

The Gaudin Hamiltonians are defined as the residue of $\tau (u|z)$ at the
point $u=z_{a}$. This results in $N$ non-local Hamiltonians%
\begin{eqnarray}
G_{a} &=&\sum_{b\neq a}^{N}\frac{1}{\sinh (2z_{a}-2z_{b})}\left\{ \frac{2}{3}%
+\cosh (2z_{a}-2z_{b})H_{a}\otimes H_{b}\right.  \nonumber \\
&&+\frac{1}{3}Y_{a}\otimes Y_{b}+2\left( S_{a}^{+}\otimes
S_{b}^{-}+S_{a}^{-}\otimes S_{b}^{+}\right) +W_{a}^{-}\otimes
W_{b}^{+}+W_{a}^{+}\otimes W_{b}^{-}  \nonumber \\
&&+\left. {\rm e}^{2(z_{a}-z_{b})}U_{a}^{+}\otimes U_{b}^{-}+{\rm e}%
^{-2(z_{a}-z_{b})}U_{a}^{-}\otimes U_{b}^{+}\right\} ,  \nonumber \\
a &=&1,2,...,N.  \label{gau.9}
\end{eqnarray}%
satisfying 
\begin{equation}
\sum_{a=1}^{N}G_{a}=0,\quad \frac{\partial G_{a}}{\partial z_{b}}=\frac{%
\partial G_{b}}{\partial z_{a}},\quad \left[ G_{a},G_{b}\right] =0,\qquad
\forall a,b.  \label{gau.10}
\end{equation}

In the next section we will use the data of the Bethe Ansatz presented in
the previous section in order to find the exact spectrum and eigenvectors
for each of these $N-1$ independents Hamiltonians.

We complete this section deriving \ the $A_{2}^{(2)}$ Gaudin algebra from
the semiclassical limit of the fundamental commutation relation (\ref{gra.5}%
): the semiclassical expansions of $T$ and $R$ can be written in the
following form 
\begin{equation}
T(u|z)=1+2\eta \lbrack l(u|z)+\beta (u|z)]+o(\eta ^{2}),\quad R(u)=P\left[
1+2\eta \lbrack r(u)+\beta (u)]+o(\eta ^{2})\right] .  \label{gau.11}
\end{equation}%
where%
\begin{equation}
\beta (u|z)=\frac{2}{3}\sum_{a=1}^{N}\frac{1}{\sinh (2u-2z_{a})}.
\label{gau.11b}
\end{equation}

Using (\ref{gau.3}--\ref{gau.4}) one can see that the \ \textquotedblright
classical $l$-operator \textquotedblright\ has the form%
\begin{equation}
l(u|z)=\left( 
\begin{array}{ccccc}
{\cal Y}(u|z)+{\cal H}(u|z) &  & {\cal W}^{-}(u|z)+{\cal U}^{-}(u|z) &  & 
{\cal S}^{-}(u|z) \\ 
&  &  &  &  \\ 
{\cal W}^{+}(u|z)+U^{+}(u|z) &  & -2{\cal Y}(u|z) &  & -{\cal W}^{-}(u|z)+%
{\cal U}^{-}(u|z) \\ 
&  &  &  &  \\ 
{\cal S}^{+}(u|z) &  & -{\cal W}^{+}(u|z)+{\cal U}^{+}(u|z) &  & {\cal Y}%
(u|z)-{\cal H}(u|z)%
\end{array}%
\right)  \label{gau.12}
\end{equation}%
where 
\begin{eqnarray}
{\cal H}(u|z) &=&\sum_{a=1}^{N}\coth (2u-2z_{a})H_{a},\qquad {\cal Y}%
(u|z)=\sum_{a=1}^{N}\frac{\frac{1}{3}Y_{a}}{\sinh (2u-2z_{a})}  \nonumber \\
{\cal W}^{\mp }(u|z) &=&\sum_{a=1}^{N}\frac{W_{a}^{\mp }}{\sinh (2u-2z_{a})}%
,\qquad {\cal U}^{\mp }(u|z)=\sum_{a=1}^{N}\frac{{\rm e}^{\pm 2(u-z_{a})}}{%
\sinh (2u-2z_{a})}U_{a}^{\mp }  \nonumber \\
{\cal S}^{\mp }(u|z) &=&\sum_{a=1}^{N}\frac{2S_{a}^{\mp }}{\sinh (2u-2z_{a})}%
.  \label{gau.13}
\end{eqnarray}%
The corresponding semiclassical $r$-matrix has the form 
\begin{eqnarray}
r(u) &=&\frac{1}{\sinh 2u}\left\{ \frac{1}{3}Y\otimes Y+\cosh 2u\ H\otimes
H+2(S^{+}\otimes S^{-}+S^{-}\otimes S^{+})\right.  \nonumber \\
&&+\left. {\rm e}^{-2u}\ U^{-}\otimes U^{+}+{\rm e}^{2u}U^{+}\otimes U^{-}+\
W^{-}\otimes W^{+}+W^{-}\otimes W^{+}\right\} .  \label{gau.14}
\end{eqnarray}%
Here we notice that\ (\ref{gau.14}) is equivalent to the $r$-matrix
constructed out of the quadratic Casimir (\ref{gau.6}) in a standard way 
\cite{Ku2}.

Substituting (\ref{gau.14}) and (\ref{gau.12}) into (\ref{gra.5}), we have 
\begin{eqnarray}
&&{\cal P}l(u|z)\otimes l(v|z)+{\cal P}r(u-v)\left[ l(u|z)\otimes 1+1\otimes
l(v|z)\right]  \nonumber \\
&=&l(v|z)\otimes l(u|z){\cal P}+\left[ l(v|z)\otimes 1+1\otimes l(u|z)\right]
{\cal P}r(u-v),  \label{gau.16}
\end{eqnarray}%
whose consistence is guaranteed by the classical {\small YB} equation (\ref%
{int.2}).

>From (\ref{gau.16}) we can derive the commutation relations between the
matrix elements of $l(u|z)$. This gives us the defining relations of the $%
A_{2}^{(2)}$ Gaudin algebra :%
\begin{eqnarray}
\lbrack {\cal H}(u|z),{\cal H}(v|z)] &=&0,\quad \lbrack {\cal H}(u|z),{\cal Y%
}(v|z)]=0,  \nonumber \\
\lbrack {\cal H}(u|z),{\cal S}^{\mp }(v|z)] &=&\pm 2\frac{{\cal S}^{\mp
}(u|z)-\cosh (2u-2v){\cal S}^{\mp }(v|z)}{\sinh (2u-2v)},  \nonumber \\
\lbrack {\cal H}(u|z),{\cal W}^{\mp }(v|z)] &=&\pm \frac{{\cal W}^{\mp
}(u|z)-\cosh (2u-2v){\cal W}^{\mp }(v|z)}{\sinh (2u-2v)},  \nonumber \\
\lbrack {\cal H}(u|z),{\cal U}^{\mp }(v|z),] &=&\pm \frac{{\rm e}^{\mp
2(u-v)}{\cal U}^{\mp }(u|z)-\cosh (2u-2v){\cal U}^{\mp }(v|z)}{\sinh (2u-2v)}%
,  \nonumber \\
\quad \lbrack {\cal Y}(u|z),{\cal Y}(v|z)] &=&0,\quad \lbrack {\cal Y}(u|z),%
{\cal S}^{\pm }(v|z)]=0,  \nonumber \\
\lbrack {\cal Y}(u|z),{\cal W}^{\pm }(v|z)] &=&\mp \frac{{\cal U}^{\pm
}(u|z)-{\cal U}^{\pm }(v|z)}{\sinh (2u-2v)},  \nonumber \\
\lbrack {\cal Y}(u|z),{\cal U}^{\pm }(v|z)] &=&\mp \frac{{\rm e}^{\pm 2(u-v)}%
{\cal W}^{\pm }(u|z)-{\cal W}^{\pm }(v|z)}{\sinh (2u-2v)}  \nonumber \\
\lbrack {\cal S}^{\pm }(u|z),{\cal S}^{\pm }(v|z)] &=&0,\quad \lbrack {\cal S%
}^{\pm }(u|z),{\cal W}^{\pm }(v|z)]=0,\quad \lbrack {\cal S}^{\pm }(u|z),%
{\cal U}^{\pm }(v|z)]=0  \nonumber \\
\lbrack {\cal S}^{\pm }(u|z),{\cal S}^{\mp }(v|z)] &=&\mp 4\frac{{\cal H}%
(u|z)-{\cal H}(v|z)}{\sinh (2u-2v)},  \nonumber \\
\lbrack {\cal S}^{\pm }(u|z),{\cal U}^{\mp }(v|z)] &=&\mp 2\frac{{\rm e}%
^{\mp 2(u-v)}{\cal W}^{\pm }(u|z)-{\cal W}^{\pm }(v|z)}{\sinh (2u-2v)}, 
\nonumber \\
\lbrack {\cal S}^{\pm }(u|z),{\cal W}^{\mp }(v|z)] &=&\pm 2\frac{{\cal U}%
^{\pm }(u|z)-{\cal U}^{\pm }(v|z)}{\sinh (2u-2v)},  \nonumber \\
\lbrack {\cal U}^{\pm }(u|z),{\cal U}^{\pm }(v|z)] &=&0,\quad \lbrack {\cal W%
}^{\pm }(u|z),{\cal W}^{\pm }(v|z)]=0,  \nonumber \\
\lbrack {\cal U}^{\pm }(u|z),{\cal U}^{\mp }(v|z)] &=&\mp \frac{{\rm e}^{\mp
2(u-v)}\left( {\cal H}(u|z)-{\cal H}(v|z)\right) }{\sinh (2u-2v)},  \nonumber
\\
\lbrack {\cal U}^{\pm }(u|z),{\cal W}^{\pm }(v|z)] &=&\pm \frac{{\cal S}%
^{\pm }(u|z)-{\rm e}^{\mp 2(u-v)}{\cal S}^{\pm }(v|z)}{\sinh (2u-2v)}, 
\nonumber \\
\lbrack {\cal U}^{\pm }(u|z),{\cal W}^{\mp }(v|z)] &=&\mp 3\frac{{\cal Y}%
(u|z)-{\rm e}^{\mp 2(u-v)}{\cal Y}(v|z)}{\sinh (2u-2v)},  \nonumber \\
\lbrack {\cal W}^{\pm }(u|z),{\cal W}^{\mp }(v|z)] &=&\mp \frac{{\cal H}%
(u|z)-{\cal H}(v|z)}{\sinh (2u-2v)}.  \label{gau.17}
\end{eqnarray}%
A direct consequence of these relations is the commutativity of $\tau
^{(2)}(u|z)$ 
\begin{equation}
\lbrack \tau ^{(2)}(u|z),\tau ^{(2)}(v|z)]=0,\qquad \forall u,v
\label{gau.18}
\end{equation}%
from which the commutativity of the Gaudin Hamiltonians $G_{a}$ follows
immediately.

\section{Off-shell Gaudin Equation}

In order to get the semiclassical limit of the {\small OSBAE} (\ref{inh.4})
we first consider the semiclassical expansions of the Bethe vectors defined
in (\ref{inh.5}), (\ref{inh.7}) and (\ref{inh.9}): 
\begin{eqnarray}
\Psi _{n}(u_{1},...,u_{n}|z) &=&(2\eta )^{n}\Phi _{n}(u_{1},...,u_{n}|z)+%
{\rm o}(\eta ^{n+1}),  \nonumber \\
&&  \nonumber \\
\Psi _{(n-1)}^{j} &=&-2(2\eta )^{n+1}\frac{\left[ {\cal U}^{-}(u|z){\rm e}%
^{-2(u-u_{j})}+{\cal W}^{-}(u|z)\right] }{\sinh (2u-2u_{j})}\Phi _{n-1}(%
\overset{\wedge }{u}_{j}|z)+{\rm o}(\eta ^{n+2}),  \nonumber \\
&&  \nonumber \\
\Psi _{(n-2)}^{lj} &=&(2\eta )^{n-1}{\cal S}^{-}(u|z)\Phi _{n-2}(\overset{%
\wedge }{u}_{l},\overset{\wedge }{u}_{j}|z)+{\rm o}(\eta ^{n}),
\label{off.1}
\end{eqnarray}%
where 
\begin{eqnarray}
\Phi _{n}(u_{1},...,u_{n}|z) &=&\left[ {\cal W}^{-}(u_{1}|z)+{\cal U}%
^{-}(u_{1}|z)\right] \Phi _{n-1}(u_{2},...,u_{n}|z)  \nonumber \\
&&+{\cal S}^{-}(u_{1}|z)\sum_{j=2}^{n}\frac{{\rm e}^{u_{j}-u_{1}}}{\cosh
(u_{j}-u_{1})}\Phi _{n-2}(u_{2},\overset{\wedge }{u}_{j}\!,u_{n}|z),
\label{off.2}
\end{eqnarray}%
with $\Phi _{0}=\left\vert 0\right\rangle $ and $\Phi _{1}(u_{1}|z)=\left[ 
{\cal W}^{-}(u_{1}|z)+{\cal U}^{-}(u_{1}|z)\right] \Phi _{0}$.

The corresponding expansions of the $c$-number functions presented in the 
{\small OSBAE} are: (\ref{inh.4}) 
\begin{eqnarray}
\Lambda _{n} &=&3+2\eta \Lambda _{n}^{(1)}+4\eta ^{2}\Lambda _{n}^{(2)}+{\rm %
o}(\eta ^{3}),  \label{off.3} \\
&&  \nonumber \\
{\cal F}_{j}^{(n-1)} &=&2\eta \ f_{j}^{(n-1)}+{\rm o}(\eta ^{2}),
\label{off.4} \\
&&  \nonumber \\
{\cal F}_{lj}^{(n-2)} &=&2(2\eta )^{3}\frac{1}{\cosh (u_{l}-u_{j})}\left\{ 
\frac{{\rm e}^{u_{j}-u_{l}}f_{l}^{(n-1)}}{\sinh (2u-2u_{l})}+\frac{{\rm e}%
^{u_{l}-u_{j}}f_{j}^{(n-1)}}{\sinh (2u-2u_{j})}\right\} +{\rm o}(\eta ^{4}),
\nonumber \\
&&  \label{off.5}
\end{eqnarray}%
\ where%
\begin{equation}
\Lambda _{n}^{(1)}=\sum_{a=1}^{N}\frac{2}{\sinh (2u-2z_{a})}  \label{off.5a}
\end{equation}%
\begin{eqnarray}
\Lambda _{n}^{(2)} &=&N+\frac{3}{2}n-\frac{3}{2}\sum_{a=1}^{N}\frac{1}{\cosh
(u-z_{a})^{2}}+\frac{1}{2}\sum_{j=1}^{n}\frac{1}{\cosh (u-u_{j})^{2}} 
\nonumber \\
&&  \nonumber \\
&&-\sum_{a=1}^{N}\sum_{j=1}^{n}\left\{ \coth (u-z_{a})\coth (u-u_{j})+\tanh
(u-z_{a})\tanh (u-u_{j})\right\}  \nonumber \\
&&+\sum_{a<b}^{N}\left\{ \coth (u-z_{a})\coth (u-z_{b})+\tanh (u-z_{a})\tanh
(u-z_{b})\right\}  \nonumber \\
&&+\sum_{j<k}^{n}\left\{ \coth (u-u_{j})\coth (u-u_{k})+\tanh (u-u_{j})\tanh
(u-u_{k})\right.  \nonumber \\
&&+\left. \frac{4}{\sinh (2u-2u_{j})\sinh (2u-2u_{k})}\right\}  \label{off.6}
\end{eqnarray}%
and 
\begin{equation}
f_{j}^{(n-1)}=\sum_{a=1}^{N}\coth (u_{j}-z_{a})-\sum_{k\neq j}^{n}\left\{
2\coth (u_{j}-u_{k})-\tanh (u_{j}-u_{k})\right\} .  \label{off.7}
\end{equation}

Substituting these expressions into the Eq.(\ref{inh.4}) and comparing the
coefficients of the terms $2(2\eta )^{n+2}$ we get the first non-trivial
consequence for the semiclassical limit of the \ {\small OSBAE}: 
\begin{equation}
\tau ^{(2)}(u|z)\ \Phi _{n}(u_{1},...,u_{n}|z)=\Lambda _{n}^{(2)}\ \Phi
_{n}(u_{1},...,u_{n}|z)-\sum_{j=1}^{n}\frac{2f_{j}^{(n-1)}\Theta _{(n-1)}^{j}%
}{\sinh (2u-2u_{j})}.  \label{off.8}
\end{equation}%
Note that in this limit the contributions from $\Psi _{(n-1)}^{j}$ and $\Psi
_{(n-2)}^{lj}$ are combined to give a new vector valued function 
\begin{eqnarray}
\Theta _{(n-1)}^{j} &=&\left[ {\cal U}^{-}(u|z){\rm e}^{-2(u-u_{j})}+{\cal W}%
^{-}(u|z)\right] \ \Phi _{n-1}(\ \overset{\wedge }{u}_{j}|z)  \nonumber \\
&&+{\cal S}^{-}(u|z)\sum_{k=1,\ k\neq j}^{n}\frac{{\rm e}^{u_{k}-u_{j}}}{%
\cosh (u_{k}-u_{j})}\ \Phi _{n-2}(\ \overset{\wedge }{u}_{j},\overset{\wedge 
}{u}_{k}|z),  \label{off.9}
\end{eqnarray}

Finally, we take the residue of (\ref{off.8}) at the point $u=z_{a}$ to get
the off-shell Gaudin equation: 
\begin{eqnarray}
G_{a}\Phi _{n}(u_{1},...,u_{n}|z) &=&g_{a}\Phi
_{n}(u_{1},...,u_{n}|z)+\sum_{l=1}^{n}\frac{2f_{l}^{(n-1)}\phi _{(n-1)}^{l}}{%
\sinh (2u_{l}-2z_{a})},  \nonumber \\
a &=&1,2,...,N  \label{off.11}
\end{eqnarray}%
where $g_{a}$ is the residue of $\Lambda _{n}^{(2)}$ 
\begin{equation}
g_{a}={\rm res}_{u=z_{a}}\Lambda _{n}^{(2)}=\sum_{b\neq a}^{N}\coth
(z_{a}-z_{b})-\sum_{l=1}^{n}\coth (z_{a}-u_{l}),  \label{off.12}
\end{equation}%
and $\phi _{(n-1)}^{l}$ is the residue of $\Theta _{(n-1)}^{l}$ 
\begin{eqnarray}
\phi _{(n-1)}^{j} &=&{\rm res}_{u=z_{a}}\Theta _{(n-1)}^{j}  \nonumber \\
&=&\frac{1}{2}({\cal U}_{a}^{-}{\rm e}^{2u_{j}-2z_{a}}+{\cal W}_{a}^{-})\Phi
_{n-1}(\overset{\wedge }{u}_{j}|z)+{\cal S}_{a}^{-}\sum_{k\neq j}^{n}\frac{%
{\rm e}^{u_{k}-u_{j}}}{\cosh (u_{j}-u_{k})}\Phi _{n-2}(\overset{\wedge }{u}%
_{k},\overset{\wedge }{u}_{j}|z).  \label{off.13}
\end{eqnarray}

The equation (\ref{off.11}) allows us solve one of the main problems of the
Gaudin models, {\it i.e.}, the determination of the eigenvalues and
eigenvectors of the commuting Hamiltonians $G_{a}$ (\ref{gau.6}): $g_{a}$ is
the eigenvalue of $G_{a}$ with eigenfunction $\Phi _{n}$ provided $u_{l}$
are solutions of the following equations $f_{j}^{(n-1)}=0$, {\it i.e}.: 
\begin{equation}
\sum_{a=1}^{N}\coth (u_{j}-z_{a})=\sum_{k\neq j}^{n}\left\{ 2\coth
(u_{j}-u_{k})-\tanh (u_{j}-u_{k})\right\} ,\quad j=1,2,...,n.  \label{off.14}
\end{equation}%
Moreover, as we will see in the next section, the off-shell Gaudin equation (%
\ref{off.11}) provides solutions for the differential equations known as 
{\small KZ} equations.

\section{Knizhnik-Zamolodchickov equation}

The {\small KZ} differential equation 
\begin{equation}
\kappa \frac{\partial \Psi (z)}{\partial z_{i}}=G_{i}(z)\Psi (z),
\label{kz.1}
\end{equation}%
appeared first as a \ holonomic system of differential equations of
conformal blocks in {\small WZW} models of conformal field theory. Here $%
\Psi (z)$ is a function with values in the tensor product $V_{1}\otimes
\cdots \otimes V_{N}$ of representations of a simple Lie algebra, $\kappa
=k+g$ , where $k$ is the central charge of the associated Kac-Moody algebra,
and $g$ is the dual Coxeter number of the simple Lie algebra.

One of the remarkable properties of the {\small KZ} system is that the
coefficient functions $G_{i}(z)$ commute and that the form $\omega
=\sum_{i}G_{i}(z)dz_{i}$ is closed \cite{RV}: 
\begin{equation}
\frac{\partial G_{j}}{\partial z_{i}}=\frac{\partial G_{i}}{\partial z_{j}}%
,\qquad \left[ G_{i},G_{j}\right] =0.  \label{kz.2}
\end{equation}
Indeed, it was indicated in \cite{RV} that the equations (\ref{kz.2}) are
not just a flatness condition for the form $\omega $ but that the {\small KZ}
connection is actually a commutative family of connections.

In this section we will identify $G_{i}$ with our $A_{2}^{(2)}$ Gaudin
Hamiltonians $G_{a}$ derived in the previous section%
\begin{eqnarray}
G_{a} &=&\sum_{b\neq a}^{N}\frac{1}{\sinh (2z_{a}-2z_{b})}\left\{ \cosh
(2z_{a}-2z_{b})H_{a}\otimes H_{b}+\frac{1}{3}Y_{a}\otimes Y_{b}\right. 
\nonumber \\
&&+2\left( S_{a}^{+}\otimes S_{b}^{-}+S_{a}^{-}\otimes S_{b}^{+}\right)
+W_{a}^{-}\otimes W_{b}^{+}+W_{a}^{+}\otimes W_{b}^{-}  \nonumber \\
&&+\left. {\rm e}^{2(z_{a}-z_{b})}U_{a}^{+}\otimes U_{b}^{-}+{\rm e}%
^{-2(z_{a}-z_{b})}U_{a}^{-}\otimes U_{b}^{+}\right\} ,  \nonumber \\
a &=&1,2,...,N.
\end{eqnarray}%
and show that the corresponding differential equations (\ref{kz.1}) can be
solved via the off-shell Bethe Ansatz method.

Let us now define the vector-valued function $\Psi (z_{1},...,z_{N})$
through multiple contour integrals of the Bethe vectors (\ref{off.2}) 
\begin{equation}
\Psi (z_{1},...,z_{N})=\oint \cdots \oint {\cal X}(u|z)\Phi
_{n}(u|z)du_{1}...du_{n},  \label{kz.3}
\end{equation}
where ${\cal X}$ $(u|z)={\cal X}$ $(u_{1},...,u_{n},z_{1},...,z_{N})$ is a
scalar function which in this stage is still undefined.

We assume that $\Psi (z_{1},...,z_{N})$ is a solution of the equations 
\begin{equation}
\kappa \frac{\partial \Psi (z_{1},...,z_{N})}{\partial z_{a}}=G_{a}\Psi
(z_{1},...,z_{N}),\quad a=1,2,...,N  \label{kz.4}
\end{equation}
where $G_{a}$ are the Gaudin Hamiltonians (\ref{gau.6}) and $\kappa $ is a
constant.

Substituting (\ref{kz.3}) into (\ref{kz.4}) we have 
\begin{equation}
\kappa \frac{\partial \Psi (z_{1},...,z_{N})}{\partial z_{a}}=\oint \left\{
\kappa \frac{\partial {\cal X}(u|z)}{\partial z_{a}}\Phi _{n}(u|z)+\kappa 
{\cal X}(u|z)\frac{\partial \Phi _{n}(u|z)}{\partial z_{a}}\right\} du,
\label{kz.5}
\end{equation}
where we are using a compact notation $\oint \left\{ \circ \right\} du=\oint
\ldots \oint \left\{ \circ \right\} $\ $du_{1}\cdots du_{n}.$

Using the Gaudin algebra (\ref{gau.17}) one can derive the following
non-trivial identity 
\begin{equation}
\frac{\partial \Phi _{n}}{\partial z_{a}}=-\sum_{l=1}^{n}\frac{\partial }{%
\partial u_{l}}\left( \frac{2\phi _{(n-1)}^{l}}{\sinh (2u_{l}-2z_{a})}%
\right) ,  \label{kz.6}
\end{equation}%
which allows us write (\ref{kz.5}) in the form 
\begin{eqnarray}
\kappa \frac{\partial \Psi }{\partial z_{a}} &=&\oint \left\{ \kappa \frac{%
\partial {\cal X}(u|z)}{\partial z_{a}}\Phi _{n}(u|z)+\sum_{l=1}^{n}\kappa 
\frac{\partial {\cal X}(u|z)}{\partial u_{l}}\left( \frac{2\phi _{(n-1)}^{l}%
}{\sinh (2u_{l}-2z_{a})}\right) \right\} du  \nonumber \\
&&-\kappa \sum_{l=1}^{n}\oint \frac{\partial }{\partial u_{l}}\left( {\cal X}%
(u|z)\frac{2\phi _{(n-1)}^{l}}{\sinh (2u_{l}-2z_{a})}\right) du.
\label{kz.7}
\end{eqnarray}%
It is evident that the last term of (\ref{kz.7}) vanishes, because the
contours are closed. Moreover, if the scalar function ${\cal X}(u|z)$
satisfies the following differential equations 
\begin{equation}
\kappa \frac{\partial {\cal X}(u|z)}{\partial z_{a}}=g_{a}{\cal X}%
(u|z),\qquad \kappa \frac{\partial {\cal X}(u|z)}{\partial u_{j}}%
=f_{j}^{(n-1)}{\cal X}(u|z),  \label{kz.8}
\end{equation}%
we are recovering the off-shell Gaudin equation (\ref{off.11}) from the
first term in (\ref{kz.7}).

Taking into account the definition of the scalar functions \ $f_{j}^{(n-1)}$(%
\ref{off.7}) and $g_{a}$ (\ref{off.12}), we can see that the consistency
condition of the system (\ref{kz.8}) is insured by the zero curvature
conditions $\partial f_{j}^{(n-1)}/\partial z_{a}=\partial g_{a}/\partial
u_{j}$. Moreover, the solution of (\ref{kz.8}) is easily obtained 
\begin{equation}
{\cal X}(u|z)=\prod_{a\neq b}^{N}\sinh (z_{a}-z_{b})^{1/\kappa }\prod_{j\neq
k}^{n}\left[ \cosh (u_{j}-u_{k})^{1/\kappa }\sinh (u_{j}-u_{k})^{-2/\kappa }%
\right] \prod_{a=1}^{N}\prod_{j=1}^{n}\sinh (z_{a}-u_{j})^{-1/\kappa }.
\label{kz.9}
\end{equation}%
This function determines the monodromy of $\Psi (z_{1},...,z_{N})$ as
solution of the trigonometric {\small KZ} equation (\ref{kz.4}) and these
results are in agreement with the Schechtman-Varchenko construction for
multiple contour integral as solutions of the {\small KZ} equation in an
arbitrary simple Lie algebra \cite{SV}.

\section{Conclusion}

In this paper the $A_{2}^{(2)}$ $19$-vertex model was investigated
generalizing previous rational vertex models results relating the Gaudin
magnet models to the semiclassical off-shell algebraic Bethe Ansatz of these
vertex models.

Using the semiclassical limit of the transfer matrix of the vertex model we
derived the trigonometric $A_{2}^{(2)}$ Gaudin Hamiltonians. The reduction
of the off-shell Gaudin equation \ to an eigenvalue equation gives us the
exact spectra and eigenvectors for the corresponding Gaudin magnets. Data of
the off-shell Gaudin equation were used to show that a hypergeometric type
integral (\ref{kz.3}) solves the trigonometric {\small KZ} differential
equation. \ 

Our results corroborate the method already used with success to construct
solutions of trigonometric {\small KZ} equations \cite{B2, CH, LU, KUM} and
elliptic {\small KZ}-Bernard equations \cite{B3}.

\vspace{1cm}

{\bf Acknowledgment:} This work was supported in part by Funda\c{c}\~{a}o de
Amparo \`{a} Pesquisa do Estado de S\~{a}o Paulo--FAPESP--Brasil, by
Conselho Nacional de Desenvol\-{}vimento--CNPq--Brasil.

\end{document}